\documentclass[sigconf]{acmart}

\usepackage{bm}
\usepackage{graphicx}
\usepackage{enumerate}
\usepackage{enumitem}
\usepackage{makecell}
\usepackage{multirow}
\usepackage{stfloats}
\usepackage{amsmath}
\usepackage{color}
\usepackage{amsfonts}
\usepackage{mathtools} 
\usepackage{verbatim}
\usepackage[normalem]{ulem}
\useunder{\uline}{\ul}{}
\AtBeginDocument{%
  \providecommand\BibTeX{{%
    \normalfont B\kern-0.5em{\scshape i\kern-0.25em b}\kern-0.8em\TeX}}}

\usepackage[skip=6pt]{caption}

\setcopyright{acmcopyright}
\copyrightyear{2018}
\acmYear{2018}
\acmDOI{XXXXXXX.XXXXXXX}

\acmConference[WWW'2023]{The Web Conference}{April 30 - May 4, 2023}{Austin, Texas, USA}
\acmPrice{15.00}
\acmISBN{978-1-4503-XXXX-X/18/06}

\begin{document}

\title{Divide and Conquer: Towards Better Embedding-based Retrieval for Recommender Systems from a Multi-task Perspective}

\author{Yuan Zhang}
\affiliation{%
  \institution{Kuaishou Technology}
  \city{}\country{}
}
\email{yuanz.pku@gmail.com}
\author{Xue Dong}
\affiliation{%
  \institution{Shandong University}
  \city{}\country{}
}
\authornote{Work done during internship at Kuaishou.}
\email{dongxue.sdu@gmail.com}
\author{Weijie Ding}
\affiliation{%
  \institution{Kuaishou Technology}
  \city{}\country{}
}
\email{dingweijie@kuaishou.com}
\author{Biao Li}
\affiliation{%
  \institution{Kuaishou Technology}
  \country{}
}
\email{biaoli6@139.com}
\author{Peng Jiang}
\affiliation{%
  \institution{Kuaishou Technology}
  \city{}\country{}
}
\authornote{Corresponding author.}

\email{jp2006@139.com}
\author{Kun Gai}
\affiliation{%
  \institution{Unaffiliated}
  \city{}\country{}
}
\email{gai.kun@qq.com}

\renewcommand{\shortauthors}{Zhang and Dong, et al.}

\begin{abstract}
Embedding-based retrieval (EBR) methods are widely used in modern recommender systems thanks to its simplicity and effectiveness. However, along the journey of deploying and iterating on EBR in production, we still identify some fundamental issues in existing methods. First, when dealing with large corpus of candidate items, EBR models often have difficulties in balancing the performance on distinguishing highly relevant items (positives) from both irrelevant ones (easy negatives) and from somewhat related yet not competitive ones (hard negatives). Also, we have little control in the diversity and fairness of the retrieval results because of the ``greedy'' nature of nearest vector search. These issues compromise the performance of EBR methods in large-scale industrial scenarios.

This paper introduces a simple and proven-in-production solution to overcome these issues. The proposed solution takes a divide-and-conquer approach: the whole set of candidate items are divided into multiple clusters and we run EBR to retrieve relevant candidates from each cluster in parallel; top candidates from each cluster are then combined by some controllable merging strategies. This approach allows our EBR models to only concentrate on discriminating positives from mostly hard negatives. It also enables further improvement from a multi-tasking learning (MTL) perspective: retrieval problems within each cluster can be regarded as individual tasks; inspired by recent successes in prompting and prefix-tuning, we propose an efficient task adaption technique further boosting the retrieval performance within each cluster with negligible overheads.

Both offline evaluation and live A/B experiments demonstrates the effectiveness of the proposed solution. The presented solution has already been deployed in Kuaishou (one of the most popular short-video streaming platforms in China with hundreds of millions of active users) for over four months since the positive A/B test.

\end{abstract}





\maketitle

\section{Introduction}

Finding precisely relevant items from an extremely large corpus of items in a limited time is one of the major challenges in many industrial recommender systems.
The most widely used solution is so called the two-stage approach~\cite{covington2016deep} where a \textit{candidate generation phase} first takes care of narrowing down the candidate set of items, so that more accurate but time-consuming models can be exploit in a subsequent \textit{ranking phase}.
Traditionally, candidate generation is often implemented by rule-based methods such as tag-based recommendation and item-based collaborative filtering~\cite{linden2003amazon,yang2020large}.
As deep learning has been successfully applied, embedding-based retrieval (EBR) methods~\cite{covington2016deep, yi2019sampling, yang2020mixed, huang2020embedding, liu2021que2search} are becoming prevalent today. 
In EBR, user and item features are encoded into embeddings by two parallel deep neural networks, and the distance (e.g., dot product) between user and item vectors are learned to discriminate relevant items (i.e. positives) from irrelevant ones (i.e. negatives). 
During employment, item embeddings can be pre-computed and indexed by an ANN (approximate nearest neighbor) search system (e.g. FAISS~\cite{johnson2019billion}), so that we can retrieve top-k relevant items efficiently in sub-linear time at serving.

\begin{figure}[t]
	\centering
	\includegraphics[width=\linewidth]{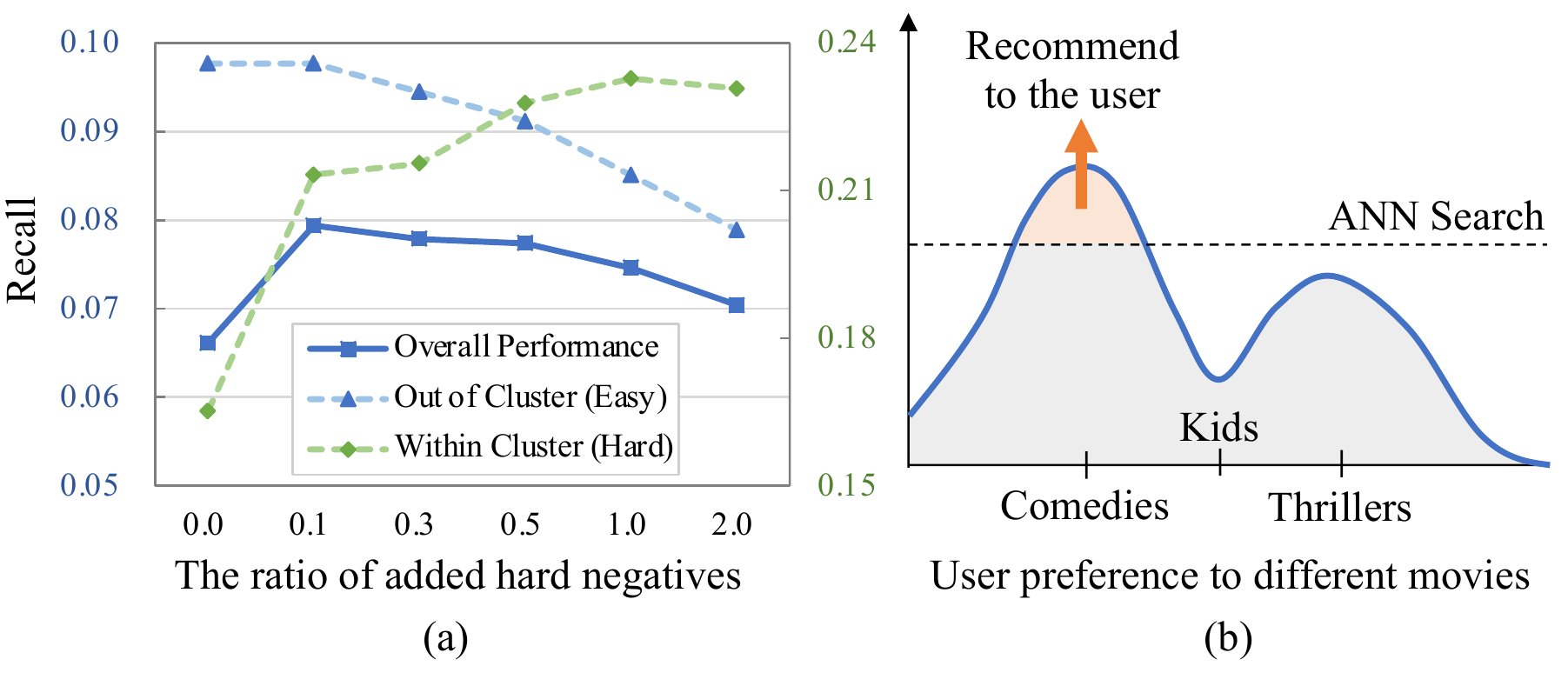}
	\caption{Motivations. (a) The retrieval performance when training EBR model with different ratios of added hard negatives (negative items from the same cluster as the positives) to easy negatives. Within Cluster and Out of Cluster indicate the performance of retrieving the ground-truth item when it is mixed with all the items from the same cluster and from other clusters, respectively. (b) Multi-modal structure of users' preferences.}
	\label{fig_motivation}
	\vspace{-0.2cm}
\end{figure}

However, despite its appealing balance between simplicity and effectiveness, EBR still faces some fundamental problems:
\begin{itemize}[leftmargin=6mm]
    \item Trade-off between discriminating easy and hard negatives. Different from the ranking phase where candidates items are mostly relevant to some degree, the candidate generation phase faces the whole corpus of items and needs to distinguish competitive candidates (positives) both from irrelevant ones (easy negatives) and from somewhat related yet not competitive ones (hard negatives). The challenging part lies in that the performance of embedding models on these two tasks are often conflicting to each other (see Figure~\ref{fig_motivation}(a)). Typically, the best practice of training EBR models is to use a mixture of easy and hard negatives with a carefully tuned mixing ratio. Nonetheless, we argue that this ostensible ``sweet spot'' is not optimal, but rather a compromise, which limits the potential of EBR models.
    
    
    \item Uncontrollable diversity and fairness~\cite{morik2020controlling} of the retrieval results.
    In most platforms, users usually have a diverse set of interested items with varying intensity, for example, who spends most of their time watching comedies may also watch thrillers sometimes. Ideally, recommender systems are supposed to present diversified results that cover all of users' information needs. However, ANN search algorithms used in EBR only greedily retrieve the top-scored items and are ignorant of the potential multi-modal nature of user intent (see Figure~\ref{fig_motivation}(b)).
    Consequently, the diversity of results completely depends on the embedding model and long-tail user interests (thrillers in the example) are prone to be underrepresented in this regard.
\end{itemize}

Motivated by these issues, we propose a simple and practical approach that be used to substantially boost the performance of EBR models almost for free.
The proposed approach is similar to a divide-and-conquer algorithm. 
More specifically, we divide the whole set of items into several clusters and assume that items within the same clusters are mostly relevant to each other. 
Instead of using an EBR model to search for relevant candidates from the whole, we run EBR on each cluster (or a subset of relevant clusters) of items in parallel and then merge results with some strategies\footnote{For example, if questing for absolute fairness among clusters, we can extract equally-sized items from each cluster as the final result.}.
This way rules out easy negatives from other irrelevant clusters and allows EBR models to only concentrate on mostly hard negatives within each cluster.
Therefore, it achieves a win-win situation where embedding models can be more accurate, while diversity and fairness become controllable by designing adequate merging strategies. 

Notably, the proposed divide-and-conquer approach naturally induces a multi-tasking learning (MTL)~\cite{crawshaw2020multi} scenario. Retrieving relevant candidates from each cluster can be regarded as an individual task. This multi-tasking perspective opens up a new direction to optimize our model accuracy.
However, in practice we find that the state-of-the-art MTL methods (e.g. MMoE~\cite{ma2018modeling}, PLE~\cite{TangLZG20}) significantly increases the computational costs especially at training, yet still not achieving as much performance gain as expected. In this paper, we draw inspirations from recent advances in prompt-based lightweight tuning~\cite{li-liang-2021-prefix, lester-etal-2021-power} to facilitate task adaptation from the input layers by introducing very few extra parameters and negligible computational overheads.
The resulting method enables much more efficient fine-tuning on each task with a unified model architecture and achieves considerably improved accuracy. With this proposed MTL paradigm, we push the potential of EBR even further.

We deployed the proposed approach on a large-scale recommender system in Kuaishou, one of the most popular short-video streaming platforms in China. Online A/B testing shows that our approach significantly improves key user engagement metrics including App Usage Time. To ensure reproducibility for a broader research community, we also conducted extensive experiments on publicly available datasets. Results show that our approach achieve up to 40\% improvements in Recall. A detailed ablation study further verifies the effectiveness of two major contributions of this paper, the proposed divide-and-conquer approach (Section~\ref{divide-n-conquer}) and the prompt-based task adaption technique (Section~\ref{prompt}).

Through this preliminary research, we hope to demonstrate a new possibility and perspective to the improvement of EBR in real-world applications. Several detail designs can be left as open questions for future studies. For instance, there might exist better and more principled ways to divide the candidate space and merge results from each cluster given different demands. Moreover, one can also explore more advanced MTL approaches for this scenario to strike a better balance between effectiveness and efficacy.

\section{Methodology}

\subsection{Background: Embedding-based Retrieval}
Suppose we have a set of users $\mathcal{U}$ and a set of items $\mathcal{I}$. 
We denote the user-side features as $\bm{x_u}$ for each user $u \in \mathcal{U}$ and item-side features as $\bm{x_i}$ for each item $i \in \mathcal{I}$. 
Embedding-based retrieval (EBR) methods use a user encoder~$f$ and an item encoder~$g$ to transform user and item features into embeddings~$\bm{e}_u = f(\bm{x_u})$ and $\bm{e}_i = g(\bm{x_i})$, respectively.
The distance (most commonly, inner product) between user embedding $\bm{e}_u$ and item embedding $\bm{e}_i$ is used to capture the relevance of item $i$ to user $u$ denoted by $r_{ui}$.

During employment, item embeddings $\{\bm{e}_i\}_{i\in \mathcal{I}}$ are calculated beforehand in an offline or near-line system and indexed by some approximate nearest neighbor (ANN) search system (e.g. FAISS~\cite{johnson2019billion}). Therefore, only the user encoder $\bm{e_u} = f(\bm{x_u})$ needs to be evaluated in real time and the ANN system can efficiently retrieve the nearest items in sub-linear time.
The most appealing part of EBR is that $f$ and $g$ can be almost any neural network models and the deployment cost is remarkably low. 

Throughout this paper, we take SASRec~\cite{KangM18} as an example to introduce the proposed solution. SASRec uses Transformer~\cite{vaswani2017attention} as the user encoder 
\begin{align}
    \label{user_encoder}
    \bm{e_u} = f(\bm{x_u}) = \text{Transformer}([\bm{s_1^{u}}, \bm{s_2^{u}}, ..., \bm{s_n^{u}}]),
\end{align}
where $\bm{s_1^{u}}, \bm{s_2^{u}}, ..., \bm{s_n^{u}}$ are the embeddings of the most recent $n$ clicked items for user $u$, and uses the latent item ID embedding alone as the item representation.

During training, the item relevance scores are trained to discriminate between users' clicked items (namely $\mathcal{I}^u$) and the rest (i.e. $\mathcal{I} \setminus \mathcal{I}^u$). For example, SASRec uses the binary cross entropy loss to train the relevance model,
\begin{equation}
\label{loss_global_neg}
    \mathcal{L} = - \sum_{u \in \mathcal{U}}\sum_{i^{+} \in \mathcal{I}^u} \Bigl[\log(\sigma(r_{ui^+})) + \mathop{\mathbb{E}}_{i^-\sim \mathcal{I} \setminus \mathcal{I}^u}\bigl[\log(1-\sigma(r_{ui^-}))\bigr]\Bigr].
\end{equation}

\subsection{Proposed Divide-and-Conquer Approach}
\label{divide-n-conquer}
As mentioned in the Introduction, the structure of negative items $\mathcal{I} \setminus \mathcal{I}^u$ can be rather diverse, where some are clearly irrelevant while some might be related but not competitive enough when compared with some others.  As the scale of candidate set $\mathcal{I}$ grows, this issue becomes more pronounced.

\begin{figure}[t]
	\centering
	\includegraphics[width=1.1\linewidth]{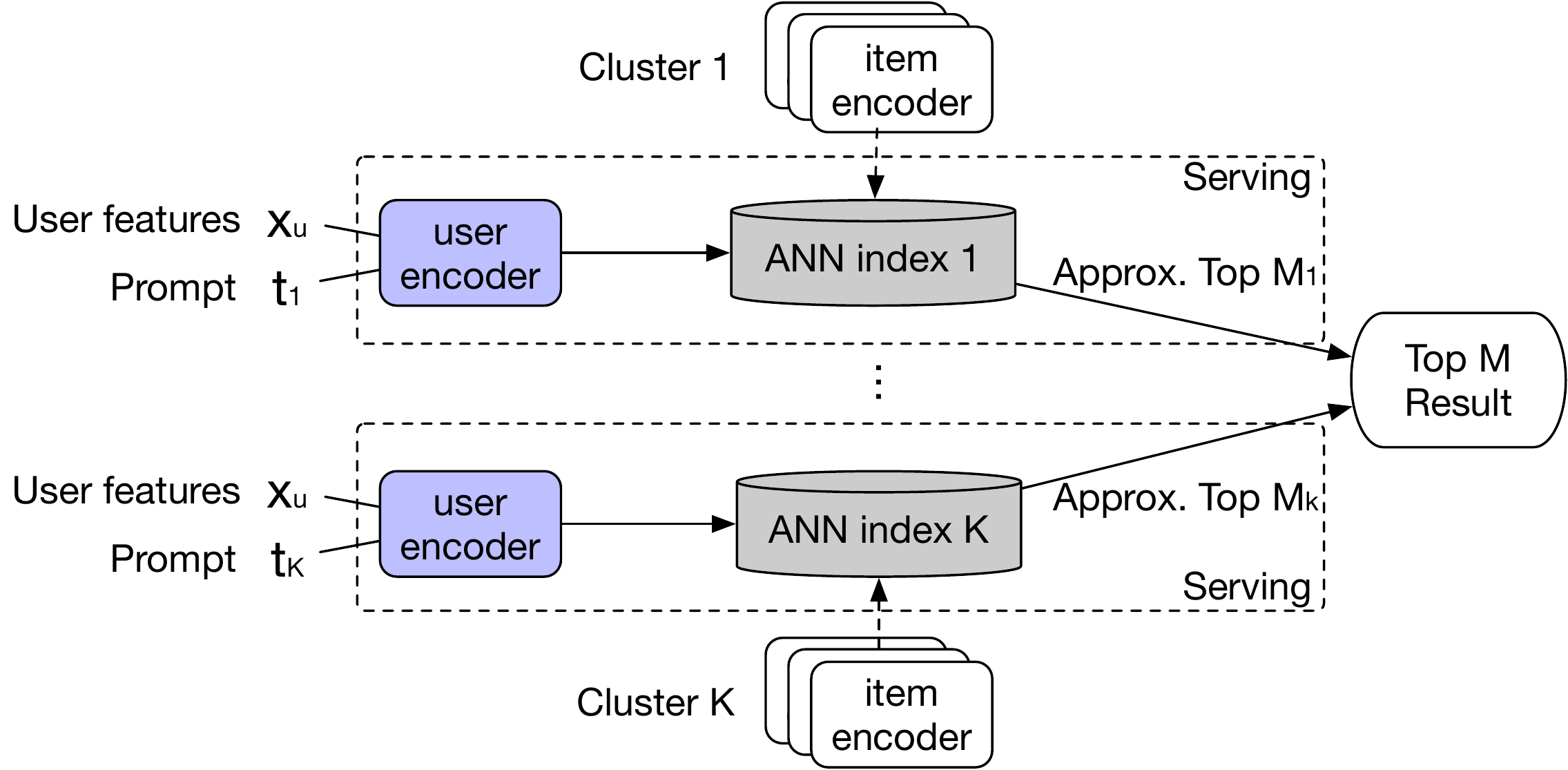}
	\caption{An overview of the system architecture.}
	\label{fig_overview}
	\vspace{-0.2cm}
\end{figure}

Our solution is motivated by the classic idea of ``divide and conquer.'' We first partition the whole candidate set into $K$ semantically relevant clusters $\mathcal{I} = \{\mathcal{C}_1, \mathcal{C}_2, ... , \mathcal{C}_K\}$ and then exploit EBR models to retrieve relevant items within each cluster instead of the whole corpus (see Figure~\ref{fig_overview}).
Note that clustering criteria can be chosen according to different application scenarios. Our offline experiments show that K-means~\cite{lloyd1982least} clustering of item embeddings trained by Word2Vec~\cite{mikolov2013efficient} can be a decent default option. In our production system, we use our in-house video categories (e.g. Sports, Gourmet, Kids etc.) predicted by content features (titles, scripts and images etc.) for both cold-start issues and explainability purposes. 

The partition of retrieval space allows the EBR models to only deal with candidates within each cluster. As as result, we only need to use negative items from the same cluster as that of the positive item (that are mostly hard negatives as mentioned before) for training as well. In other words, this way helps EBR models become more focused and hence potentially more ``productive.'' The training loss is correspondingly rewritten as 
\begin{equation}
\label{loss_intag_neg}
    \mathcal{L} = - \sum_{k=1}^{K} \sum_{u \in \mathcal{U}}\sum_{i^{+} \in \mathcal{I}^u \cap \mathcal{C}_k} \Bigl[\log(\sigma(r_{ui^+})) + \mathop{\mathbb{E}}_{i^-\sim \mathcal{C}_k \setminus \mathcal{I}^u}\bigl[\log(1-\sigma(r_{ui^-}))\bigr]\Bigr].
\end{equation}

Once the top scored items from each cluster are obtained, we can assign adequate quotas to each cluster to meet customized requirements for the final retrieval results. 
For instance, suppose we want to generate a final candidate set of size $M$. We can train a user-intent model using the same features as the user encoder and predict the probability $p_{uk}$ that the relevant items to user $u$ are likely to fall into cluster $\mathcal C_k$. Then, we combine the top-$M_k$ items from each cluster as the final results, where $M_k = M \cdot [(p_{uk})^\alpha/\sum_{k'=1}^{K} (p_{uk'})^\alpha$] and $\alpha$ is a tunable hyper-parameter\footnote{Setting $\alpha = 0$ leads to absolute fairness recommendation, while $\alpha \to +\infty$ corresponds to only recommend the most relevant cluster.}. Because EBR can be run on different clusters of items in parallel and the search space of each run is approximately $K$ times smaller, this divide-and-conquer procedure does not significantly increase the response latency.



\subsection{Prompt-like Multi-task Learning}
\label{prompt}
By looking at the training loss in Eq. (\ref{loss_intag_neg}), we are actually training $K$ separate sub-tasks where the $k$-th sub-task corresponds to retrieve relevant items from $\mathcal C_k$. Thus, there might be chances to further exploit positive transfers among these sub-tasks rather than simply blending all task samples together. This multi-task learning perspective motivates us to experiment with some state-of-the-art MTL methods, e.g. MMoE~\cite{ma2018modeling}. However, as mentioned above, neither its effectiveness nor efficiency is as good as expected (Section~\ref{offline-results}). Instead, inspired by prompt-based tuning approaches~\cite{li-liang-2021-prefix, lester-etal-2021-power,he2022hyperprompt}, we attempt to feed task identifies as prompts to user encoders to facilitate task adaption in a parameter-efficient way. That is, for samples from each cluster $\mathcal C_k$,
\begin{equation}
    \bm{e_u} = f(\bm{x_u}) = \text{Transformer}([\bm{t_{k}};\bm{s_1^{u}}, \bm{s_2^{u}}, ..., \bm{s_n^{u}}]),
\end{equation}
where $\bm{t_{k}}$ is the trainable embedding of the $k$-th task. Yet, although this approach cuts down the training time, the performance is still not improved significantly. 

We hypothesize that the reason lies in that the number of layers in Transformer used in recommendation models is much smaller than that in most NLP scenarios\footnote{The SASRec paper reports that the two-layer Transformer gives the best result.}. Thus, the interactions between prompts and raw inputs cannot be properly captured by merely a couple of layers of self-attention blocks. We then propose to impose explicit feature interactions by the Hadamard product,
\begin{equation}
    \bm{e_u} = f(\bm{x_u}) = \text{Transformer}([\bm{s_1^{u}}\odot\bm{t_{k}}, \bm{s_2^{u}}\odot\bm{t_{k}}, ..., \bm{s_n^{u}}\odot\bm{t_{k}}]).
\end{equation}
This trick shares a similar idea to the treatment in HyperPrompt~\cite{he2022hyperprompt} where each input token embedding is concatenated with the prompt.

\begin{table}[t]
\caption{Statistics of datasets (after preprocessing).}
\label{dataset_statistics}
\begin{tabular}{@{}cccccc@{}}
\toprule
Dataset & \#Users & \#Items & \#Interactions & Density & \#Clusters \\ \midrule
ML-1M & 6,040 & 3,706 & 1,000,209 & 0.045 & 10 \\
KuaiRand & 25,828 & 108,025 & 20,141,835 & 0.007 & 15 \\ \bottomrule
\end{tabular}
\vspace{-0.4cm}
\end{table}
\section{Experiments}
\subsection{Offline Evaluation}
\subsubsection{Datasets}
We conduct extensive experiments on two public datasets. The statistics of the two datasets are shown in Table~\ref{dataset_statistics}.
\begin{itemize}[leftmargin=5mm]
    \item \textbf{ML-1M}~\cite{HarperK16} consists of one million anonymous ratings of more than three thousand movies made by 6,040 MovieLens users. All the movies watched by a user are regarded as relevant.
    \item \textbf{KuaiRand}~\cite{gao2022kuairand} is an publicly available dataset collected from the logs of the recommender system in Kuaishou. In our experiment, we treat clicked items as relevant items to a user and only keep those interactions from the main recommendation scenario (the ``tab'' field equals to one). We also filter out the items and users whose frequencies are less than $70$ and $10$, respectively.
\end{itemize}

\subsubsection{Evaluation Protocol and Baselines}
Following~\cite{he2017neural,KangM18}, we hold out user's last and penultimate interacted items as ground-truths for test and validation, respectively.
To simulate real-world applications, we apply each method to generate a candidate set of $M$ relevant items from all the unseen items and evaluate its recall rate.
We set $M$ to roughly $5\%$ to $10\%$ of the total number of items, i.e. $M \in [20, 50]$ for ML-1M and $M \in [500, 1000]$ for KuaiRand.

We compare our proposed divide-and-conquer approach with the following baseline models,
\begin{itemize}[leftmargin=5mm]
    \item \textbf{Matrix Factorization (MF)}~\cite{hu2008collaborative} is a classic method recommendation method. We train the MF model with the same training objective as Eq. (\ref{loss_global_neg}).
    \item \textbf{SASRec}~\cite{KangM18} adopts self-attention blocks to model user interaction sequences. We use the same hyperparameters as in their original paper unless specified otherwise.
    \item \textbf{SASRec+} is an extension of SASRec that trained with a mixture of easy negatives and hard negatives (negative items from the same cluster as the positive item). We tune the mixing ratios on the validation set and report the best results.
    \item \textbf{MIND}~\cite{LiLWXZHKCLL19} uses the routing mechanism as in capsule networks to group user behaviors into multiple clusters and obtain multiple user embeddings for retrieval.
    \item \textbf{ComiRec-SA}~\cite{CenZZZYT20} uses multi-head attention mechanisms to generate multiple embeddings for each user to capture their diverse interests.
    \item \textbf{MMoE}~\cite{ma2018modeling} is a widely used multi-task learning method. We adopt self-attention blocks as individual experts and use four experts in our experiments.
\end{itemize}

\begin{table}[t]
\caption{Performance comparison on ML-1M and KuaiRand. 
The last row reports the relative improvement of our proposed method over its base model, SASRec. R@M is short for Recall@M.}
\label{comparsion}
\begin{tabular}{@{}ccccc@{}}
\toprule
\multirow{2}{*}{Method} & \multicolumn{2}{c}{ML-1M} & \multicolumn{2}{c}{KuaiRand} \\ \cmidrule(lr){2-3} \cmidrule(lr){4-5}
 & R@20 & R@50 & R@500 & R@1000 \\ \midrule
MF & 0.120 & 0.231 & 0.086 & 0.132 \\
SASRec & 0.183 & 0.337 & 0.178 & 0.254 \\
SASRec+ & 0.220 & 0.367 & 0.200 & 0.284 \\
MIND & 0.186 & 0.343 & 0.138 & 0.201 \\
ComiRec & 0.192 & 0.341 & 0.188 & 0.269 \\ \midrule
Ours & \textbf{0.234} & \textbf{0.375} & \textbf{0.254} & \textbf{0.359} \\
w/ Naive MTL & 0.221 & 0.366 & {\ul 0.245} & {\ul 0.344} \\
w/ MMoE & {\ul 0.224} & {\ul 0.371} & 0.239 & 0.341 \\ \midrule
Improvement & +27.9\% & +11.3\% & +42.7\% & +41.3\% \\ \bottomrule
\end{tabular}
\end{table}


\subsubsection{Experimental results}
\label{offline-results}
Table~\ref{comparsion} summarizes the performance of the proposed approach as well as baselines on ML-1M and Kuairand. First, the proposed approach outperforms all the baseline methods consistently on both datasets and achieves up to $27.9\%$ and $42.7\%$ improvements over its base model SASRec on ML-1M and KuaiRand datasets, respectively. 
As mentioned in the Introduction, although SASRec+ indeed outperforms SASRec by using a carefully tuned mixture of easy and hard negatives, there is still a big gap between SASRec+ and ours.

We also did an ablation study to assess both the proposed divide-and-conquer approach and our prompt-like MTL method. Results show that by dropping the prompt module (i.e. using naive MTL with a shared SASRec model for all tasks instead), the overall performance drops by up to $5.6\%$ and $4.2\%$ on ML-1M and KuaiRand, respectively. Yet, it still significantly improves the performance of SASRec+, which demonstrates that our divide-and-conquer strategy does work as expected.
To prove it further, we evaluate the within-cluster retrieval performance by restricting the candidate set to those within the same cluster as the ground-truth item. As shown in 
Table~\ref{intag_comparison}, by only focusing the within-cluster negatives during training, Naive MTL can already substantially improve the retrieval performance within each cluster and then lead to the overall performance gains. When adding prompts to facilitate task adaption, the within-cluster retrieval performance performs even better. 

Meanwhile, MMoE achieves comparable results on ML-1M, but we can see from Table~\ref{throughput} that it dramatically increases the training cost and the training throughput declines by around $70\%$. In contrast, the proposed method incurs almost no extra training cost.

\begin{table}[t]
\caption{The retrieval performance, within the clusters of ground-truth items, of different MTL methods.}
\label{intag_comparison}
\begin{tabular}{@{}ccccccc@{}}
\toprule
\multirow{2}{*}{Method} & \multicolumn{3}{c}{ML-1M} & \multicolumn{3}{c}{KuaiRand} \\ 
\cmidrule(lr){2-4}\cmidrule(lr){5-7}
 & R@5 & R@20 & R@50 & R@100 & R@500 & R@1000 \\ \midrule
SASRec & 0.191 & 0.424 & 0.639 & 0.172 & 0.399 & 0.543 \\
Naive MTL & 0.217 & 0.463 & 0.674 & 0.228 & 0.502 & 0.658 \\
MMoE & 0.223 & 0.482 & \textbf{0.682} & 0.218 & 0.496 & 0.655 \\
Ours & \textbf{0.235} & \textbf{0.483} & \textbf{0.682} & \textbf{0.237} & \textbf{0.522} & \textbf{0.682} \\ \bottomrule
\end{tabular}
\vspace{-0.1cm}

\end{table}

\begin{table}[t]
\caption{The training throughput of different methods.}
\label{throughput}
\begin{tabular}{@{}cccc@{}}
\toprule
Method & SASRec & MMoE & Ours (Prompt-like) \\ \midrule
Throughput (samples/sec.) & 25K & 7.2K & 25K \\ \bottomrule
\end{tabular}
\vspace{-0.1cm}
\end{table}

\subsection{Live A/B Experiment}
The proposed approach has also been tested in a live A/B experiment in Kuaishou. The A/B test lasted for three days (July 3-5, 2022) and involved over 40 million users in the experiment group, yielding strongly statistically significant experimental results.
In the experiment group, the proposed retrieve method serves as one of the candidate sources in the candidate generation phase. Compared with the control group, our method leads to significant improvements on most of the key user engagement metrics as shown in Table \ref{ab-test}.
We further compare the engagement rates of items eventually shown to users from our source with those from ComiRec. As shown in Table \ref{online-rate}, our approach performs better as well.

\begin{table}[t]
\caption{Results of the live A/B experiment in Kuaishou. All the performance gains are statistically significant at $p < 0.05$.}
\label{ab-test}
\begin{tabular}{@{}ccccc@{}}
\toprule
App Usage Time & Likes & Follows & Shares & Downloads \\ \midrule
+0.096\% & +0.75\% & +1.01\% & +1.04\% & +2.40\% \\ \bottomrule
\end{tabular}
\vspace{-0.1cm}
\end{table}

\begin{table}[t]
\caption{Comparison between online engagement rates of the recommended items from ComiRec and our approach.}
\label{online-rate}
\begin{tabular}{@{}ccccc@{}}
\toprule
Metrics (\%) & Click Rate & Like Rate & Follow Rate & Share Rate  \\ \midrule
ComiRec & 56.1 & 2.85 & 0.283 & 0.209 \\
Ours & 60.5 & 2.92 & 0.431 & 0.239 \\
Improvement & +7.8\% & +2.5\% & +52.3\% & +14.4\% \\ \bottomrule
\end{tabular}
\vspace{-0.3cm}
\end{table}

\section{Conclusion}
This paper introduces a simple and production-proven solution to overcome existing limitations of embedding-based retrieval methods and effectively boost their performance. Our contribution is twofold, the divide-and-conquer approach (Section~\ref{divide-n-conquer}) and the prompt-like MTL technique (Section~\ref{prompt}). Extensive results in both offline and live A/B experiments demonstrate the effectiveness of the proposed solution.
\begin{acks}
We would like to thank the anonymous reviewers for their valuable comments and suggestions.
\end{acks}

\balance
\bibliographystyle{ACM-Reference-Format}
\bibliography{main}










\end{document}